\documentstyle[prd,aps,epsf,12pt]{revtex}

\newcommand{\la}{\langle}
\newcommand{\ra}{\rangle}

\newcommand{\beq}{\begin{eqnarray}}
\newcommand{\eeq}{\end{eqnarray}}
\newcommand{\delslash}{\partial \hspace{-6pt}/}

\newcommand{\sbeq}{\begin{subeqnarray}}
\newcommand{\seeq}{\end{subeqnarray}}

\newcommand{\eps}{\epsilon}
\newcommand{\Sigpi}{\Sigma_{\pi N}}
\newcommand{\cl}{\centerline}
\newcommand{\btem}{\bibitem}

\newcommand{\pipi}{{$\pi$-$\pi$ \ }}
\newcommand{\chis}{chiral symmetry }

\tighten

\begin{document}


\title{The $\sigma$-meson and  $\pi$-$\pi$ Correlation 
 in Hot/Dense Medium: \\
 soft modes for chiral transition in QCD}

\author{ T. Hatsuda$^{(1)}$ and  T. Kunihiro$^{(2)}$}
\address{$^{(1)}$  Department of Physics, University of Tokyo,
  Tokyo 113-0077 Japan}
\address{$^{(2)}$Yukawa  Institute for Theoretical Physics,
Kyoto University, Kyoto 606-8502, Japan}

\date{\today}

\maketitle

\begin{abstract}

After a brief overview of the realization of chiral symmetry
 in hot and/or dense medium, the implications of the 
 existence of a light scalar-isoscalar meson (which we call
 $\sigma$ throughout this report) are discussed
 in the vacuum and in the medium.
 Special emphasis is put on its relation to 
 the fluctuation of the chiral order parameter 
 $\langle \bar{q}q \rangle$. In the vacuum,
 the $\sigma$-meson is an elusive resonance corresponding to 
 a pole deep in the second Riemann sheet 
 of the  $\pi$-$\pi$ scattering matrix in the $I$=$J$=0 channel.
 This is because the 
 the amplitude fluctuation of  $\langle \bar{q}q \rangle$
 (corresponding to  {\rm ur}-$\sigma$)
  and the phase fluctuation of  $\langle \bar{q}q \rangle$
 (2 pion states) mix strongly in the vacuum.
 As the temperature and/or density of the 
 system increase, however, there arises a softening of 
 this complex pole due to the partial restoration of chiral symmetry.
 Such soft modes  play a key role to understand the 
 QCD phase structure.
 We demonstrate that even a slight softening of the $\sigma$  
 mode could induce strong spectral enhancement 
 and strong $\pi$-$\pi$ correlation 
 near the 2$m_{\pi}$ threshold in the  $I$=$J$=0 channel,
 which thereby can be a signature of the
 partial restoration of chiral symmetry.
 Such spectral enhancement may be seen
 not only in hot matter created by the relativistic heavy ion collisions
 but also in cold matter (heavy nuclei) probed by photons and hadrons.
 Relevance of the partial restoration 
 of chiral symmetry in heavy nuclei with the recent
 data by the CHAOS and CB collaborations as well as 
 with on-going and future  experiments are also discussed.

\end{abstract}

\newpage

{\cl {\bf CONTENTS}}

\begin{itemize}

\item[I.] {\bf Introduction}

\item[II.] {\bf The light scalar-isoscalar meson in QCD}

  \begin{itemize}

  \item[II.A.] Chiral symmetry, analyticity, crossing symmetry and
             complex $\sigma$-pole in $\pi$-$\pi$ scattering

  \item[II.B.] Hadron phenomenology and the $\sigma$-meson

  \end{itemize}
 
\item[III.] {\bf Chiral restoration and soft mode at finite $T$ and density}

  \begin{itemize}

  \item[III.A.] Chiral condensate at low $T$

  \item[III.B.] Chiral condensate at finite baryon density

  \item[III.C.] Hadronic correlations and symmetry restoration

  \end{itemize}

\item[IV.] {\bf Scalar correlation n the medium}

  \begin{itemize}

  \item[IV.A.] Spectral enhancement in hot matter

  \item[IV.B.] Spectral enhancement in nuclear matter

  \item[IV.C.] Behavior of the $\sigma$-pole in 2nd Riemann sheet

  \end{itemize}

\item[V.] {\bf Possible experimental signature}

  \begin{itemize}

  \item[V.A.] Softening in the scalar channel

  \item[V.B.] Medium effect in other channels

  \end{itemize}

\item[VI.] {\bf Summary and concluding remarks}

\end{itemize}

\newpage

\section{Introduction}

One of the most intriguing phenomena in 
 quantum chromodynamics (QCD) is the 
dynamical breaking of chiral  symmetry (DBCS).
This explains the existence of the pion and governs
most of the low energy phenomena in hadron physics.
   DBCS is associated 
with the condensation of quark - anti-quark
pairs in the QCD vacuum, 
 $\langle \bar{q}q \rangle$, which is analogous to the condensation of 
 Cooper pairs 
 in the theory of  superconductivity \cite{NJL}.
 As the temperature ($T$) and/or the baryon density ($n_B$) increase,
 the QCD vacuum undergoes a phase transition to the chirally
 symmetric phase where $\langle \bar{q}q \rangle$ vanishes.
 Studying  the mechanism of DBCS, exploring the 
   phase structure of the QCD ground state,
    and finding possible signatures of the QCD phase transition
    are the central issues in modern hadron physics \cite{HKBR}.
 
 Various phases of QCD are characterized not only by
  $T$ and $n_B$,  but also by  
   the current masses of light quarks ($m_{u,d,s}$). 
 In the real word,  the masses of $u$ and $d$ quarks are
  much lighter than the $s$ quark \cite{LEUT01},
\begin{equation}
\frac{m_s}{m_d} = 18.9 \pm 0.8, \ \ 
\frac{m_u}{m_d} = 0.553 \pm 0.043, 
\end{equation}
where $m_s (2 {\rm GeV}) (\sim 100 $ MeV) is comparable to
 $\Lambda_{QCD}$ (the QCD scale parameter) 
 and also to the critical temperature
  (density$^{1/3}$ ) of the quark-hadron transition.
  Therefore,  studying the phase structure
 in the $m_s$-$T$-$n_B$ space by neglecting
 $m_{u,d}$ as a first approximation would give us
 a good  insight into the real world. 
  
 Shown in   Fig.1 is a possible phase structure
  projected onto the $m_s$-$T$ plane with $n_B$=$0$ (the left panel)
 and onto $m_s$-$n_B$ plane with $T$=$0$ (the right panel). 
 $m_s$=$0$ ($m_s$=$\infty $) corresponds to the limit of
  $SU_L(3) \otimes SU_R(3)$ ($SU_L(2) \otimes SU_R(2)$) chiral symmetry.

 \begin{figure}[tbp]
   \centering
   \epsfxsize=16.0cm
   \epsfbox{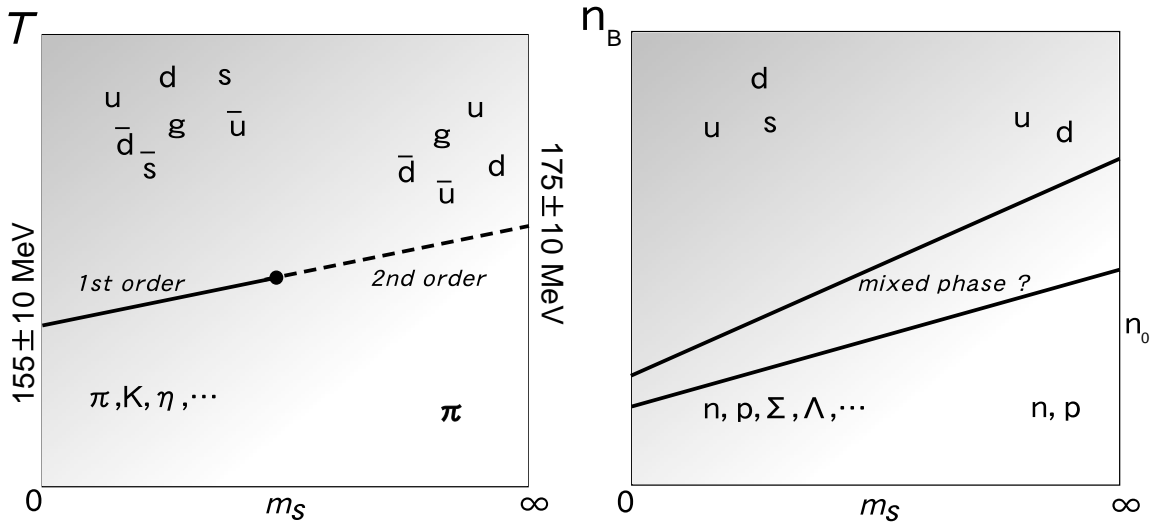}
 \vspace{0.5cm}
\caption{Possible phases in the $m_s$-$T$ plane at $n_B$=$0$ 
(left panel) and the $m_s$-$n_B$ plane at $T$=$0$. $m_{u,d}$=$0$ is assumed.
 Relevant degrees of freedom in each phase are also shown. 
$n_0 = 0.17{\rm fm}^{-3}$  is the normal nuclear matter density.}
    \label{fig1}
\end{figure}

The  chiral phase transition for $T \neq 0$ and  $n_B=0$  
 has been extensively studied on the lattice for recent years
 with $\langle \bar{u}u \rangle (= \langle \bar{d}d \rangle ) $ as
 the chiral order parameter \cite{KAR01}. 
The  analysis based on the 
renormalization group and the universality hypothesis
indicates that the chiral transition at finite $T$ is  
of second (first) order
for large (small) $m_s$ \cite{PW84}. This has been confirmed
by the lattice QCD simulations with dynamical quarks.
 The critical temperature for massless 2 and 3 flavors
has been also studied on the lattice and the results are summarized 
as
\begin{eqnarray}
    T_{pc} (N_f=3) &=& (154 \pm 8) {\rm MeV} \ \ ({\rm staggered \ fermion}), 
  \nonumber \\
    T_{pc} (N_f=2) &=& (173 \pm 8) {\rm MeV} \ \ ({\rm staggered \ fermion}),
 \\
    T_{pc} (N_f=2) &=&(171 \pm 4) {\rm MeV} \ \ ({\rm  Wilson\ fermion}), 
  \nonumber
\end{eqnarray}
where $T_{pc}$ is a pseudo-critical temperature extracted from the
 chiral susceptibility.
      
The second order line and the first order line 
 in Fig.\ref{fig1} (left panel)  is expected
to meet at some   intermediate value of $m_s$
at the so-called  tricritical point \cite{WIL92}.  
However, the precise location of this tricritical  point 
in $m_s$-$T$ plane is not identified yet.
If we have finite but small $u, d$ quark masses,  
the second order line turns into a smooth crossover,
and the tricritical point becomes an end point
of the first order line.
Therefore, in the real world where 
two light quarks + one medium-heavy quark exist,
the chiral  transition
at $T \neq 0$ is either the first order or the crossover.

The chiral phase transition at finite baryon density is not
well understood. One of the reasons is 
the first principle lattice QCD simulation
for finite chemical potential ($\mu$)  has a 
severe sign problem originating from the 
complex fermionic determinant \cite{KAR01}.
Analyses based on simple models such as   
the Nambu-Jona-Lasinio (NJL) model
show that the chiral transition may be first
order for $n_B \neq 0$ and  $T \ll T_c$ \cite{AY89}.
\footnote{The possible vector coupling between quarks 
 in the NJL model may alter the
order of the chiral transition at finite density with low
temperatures.\cite{klimt}}
If this is the case,   the quark-hadron    
mixed phase is expected in the $m_s$-$n_B$ plane as schematically
shown in the right panel of Fig.1.
The system at finite baryon density, which 
is intrinsically quantum,  is obviously  richer in physics
than that at finite $T$ and has 
various phases such as the baryon superfluidity,
meson condensations, and  the color superconductivity
(see the reviews, \cite{TAMA93}).
 
For sufficiently large $m_s$ with $m_{u,d}=0$, there will be also a 
tricritical point in the $T$-$\mu$ plane.
For small but finite  $m_{u,d}$, the tricritical point again becomes
an end point of the first order line.
Very recently, a new attempt on the lattice has been made to 
locate this  end point   by looking for 
the Lee-Yang zero of the partition function numerically,
 where the 2-dimensional reweighting method is employed
 to obtain the partition function at finite $T$ and  $\mu$  \cite{FK01}. 
 It is found that $T_{\rm end}$=(160$\pm$3.5) MeV and
 $\mu_{\rm end}$=(725$\pm$35) MeV for (2+1)-flavors
 using  the data on $4^3\times 4, 6^3 \times 4$ and 
 $8^3 \times 4$ lattices. Although obtaining a clear 
 signal of  Lee-Yang zeros may require  much larger lattice size,
 this approach provides a new way of analyzing the physics near the 
critical line in the $T$-$\mu$ plane.

\section{The light scalar-isoscalar meson in QCD}

The order parameter $\langle \bar{q}q \rangle $ of the chiral 
transition is determined as the value  
at which the effective  potential (free energy) 
$ V (\sigma)$ takes the minimum.
The light scalar-isoscalar meson, which we call 
$\sigma$ in the following,
is a particle representing the amplitude fluctuation
 of the order parameter around the minimum of $ V (\sigma)$.
In this sense, the $\sigma$ meson 
 is analogous to the Higgs particle $H$  in the 
 Glashow-Salam-Weinberg  model.
However, the properties of $\sigma$ and $H$ or,
 more precisely, the physical modes 
in the respective channels, could be quite different
 because of the following reason:
 The  Nambu-Goldstone (NG) boson, which is 
 a phase fluctuation of the order parameter,
 appears as the pion in QCD, while the NG boson
 in the electro-weak theory is absorbed into
 the gauge bosons. 
 Therefore,  $\sigma$ can have a strong width
 decaying into two pions, while such process for $H$ 
  does not exist. In other words,
 the physical $\sigma$ state  in QCD 
 can be at most a broad resonance and is represented
 by a linear superposition
 of the {\rm ur}-$\sigma$ (genuine amplitude-fluctuation of the 
order parameter) and the two-pion state.
 
\subsection{Chiral symmetry, analyticity, crossing symmetry \\
 and complex $\sigma$-pole in $\pi$-$\pi$ scattering}
 
As mentioned above, a tricky point with  the $\sigma$ meson is that 
it  couples to two pions to acquire  a large width 
$\Gamma_{\sigma}  \sim {\rm Re}\  m_{\sigma}$ and make itself 
elusive \cite{SIGMA01}.
 Nevertheless, 
 recent careful phase shift analyses of the \pipi scattering in 
the $I$=$J$=0 channel have come to 
claim a $\sigma$ pole  
in the complex energy plane  
with the real part Re $m_{\sigma}= 500$-800 MeV and the imaginary 
part Im $m_{\sigma}\simeq 500$MeV\cite{PDG00,SIGMA01,pipi}.

\begin{figure}[tbp]
   \centering
   \epsfxsize=11cm
   \epsfbox{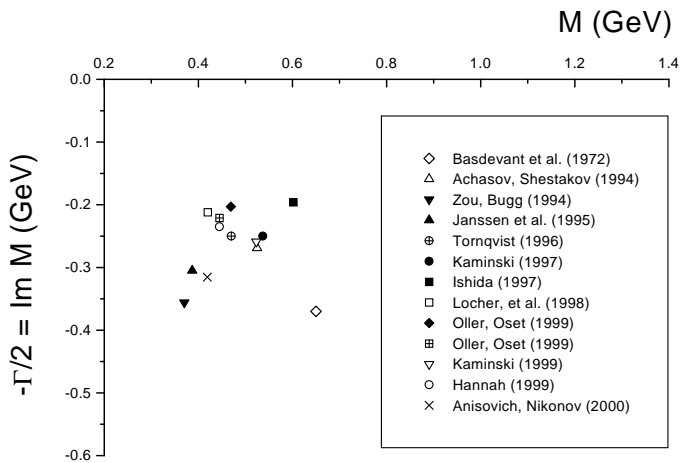}
 \vspace{0.5cm}
\caption{The $\sigma$-pole of the $\pi$-$\pi$ scattering matrix
 in the complex energy plane in various analyses.
 This figure is taken from {\protect \cite{XZ01}.}}
    \label{fig2}
\end{figure}

One of the crucial points to deduce the poles in the
 $\pi$-$\pi$ scattering matrix
 is to construct the invariant amplitude
 so that it satisfies the chiral symmetry constraints,
 the  analyticity, the unitarity and 
the (approximate) crossing symmetry.
 For this purpose, some reliable resummation method such as the 
$N/D$ method
 should be employed. This point has been  emphasized
 e.g.  in \cite{IH99} and in \cite{OOP98,CGL01}.
 In the construction of \cite{IH99},
  the ur-$\sigma$ strongly coupled to 2$\pi$
 is favored to
 have a good overall agreement with the phase shifts in
 $I$=$J$=0, $I$=$J$=1 and ($I$=2, $J$=0) phase shifts.
 In \cite{OOP98,CGL01},  the unitalization of the 
 strong $2\pi$ correlation in the $I$=$J$=0 channel is enough to
 reproduce the phase shifts.
 Although it is not clear at the moment whether one 
 really needs to introduce the ur-$\sigma$ or not to
 reproduce the phase shifts, the
 existence of the light  $\sigma$ pole with a large imaginary
part in 
 the 2nd Riemann sheet is the common feature in all 
 modern analyses.
 This is seen in Fig.\ref{fig2} where a recent summary
 (taken from \cite{XZ01}) of the complex  $\sigma$ pole is shown. 

\subsection{Hadron phenomenology and the $\sigma$ meson} 

If the location of the $\sigma$ pole
 in the 2nd Riemann sheet has Re $m_{\sigma} 
\sim 500 $ MeV, many experimental facts  which otherwise are 
mysterious  can be nicely accounted for  in a simple way.
 Some examples in hadron phenomenology where  the scalar-isoscalar
 fluctuations are relevant include the following.
 More extensive discussions can be see in \cite{SIGMA01}.

\begin{itemize}

\item 
 The scalar-isoscalar resonance with 
  Re $m_{\sigma} \sim m_{K} $ may have a sizable contribution to 
  the enhancement of the $\Delta I=1/2$ process in 
 K$ \rightarrow \pi \pi$ decay \cite{MLS}.

\item
The state-independent attraction, which is responsible for
the  nuclear binding, originates from
 the two-pion exchange potential.
 The two-pion exchange includes the {\em ladder},  
the {\em cross} and the {\em rescattering} diagrams
with the $\Delta (1232)$ being incorporated in the intermediate
states. The $\sigma$ plays a role through 
 the rescattering contribution as well as through the
 possible direct $\sigma$-$N$ coupling. The former
 should be constructed consistently with the 
\pipi phase shift in the $I$=$J$=0 channel.
For the latter, systematic analysis of 
 $N$-$N$, hyperon-$N$
and hyperon-hyperon interactions may be useful \cite{TAMA01}.

\item The $\pi$-$N$ sigma term and the generalized 
 scalar-charge  of baryons can be defined as
$\Sigpi =\hat {m}\la N |\bar{u} u + \bar {d} d | N\ra $ and
$\la B\vert \bar {q} q\vert B\ra $ ($q= u, d, s, ...$), 
respectively.  
Effective charges are usually enhanced (suppressed) due to 
collective excitations generated by the attractive (repulsive)
forces. 
The enhancement is caused by the polarization of the vacuum in the
 $I$=$J$=0 channel and can indeed account for the empirical 
value of $\Sigpi \simeq 45$ MeV without introducing a large OZI 
violation in the nucleon \cite{KH90}.

\item There are some effective theories of QCD which predict
the light {\rm ur}-$\sigma$ such as the Schwinger-Dyson approach with
 the rainbow-ladder approximation \cite{ES84}, the 
 Nambu-Jona-Lasinio model \cite{HK94}, and
 an approach based on the mended symmetry \cite{WEIN90}.

\end{itemize}
 
The above list implies that the 
correlation in the scalar-isoscalar channel
 is indeed important in the hadron dynamics. 
This is in a sense
natural because the dynamics which is responsible for the 
correlations
in this channel is nothing but the one which drives the chiral
symmetry breaking.

A fundamental question is, then, 
whether the complex $\sigma$-pole observed in the $\pi$-$\pi$ phase shift
 can be interpreted as  a quantum fluctuation of the
 order parameter associated with the dynamical breaking
 of chiral symmetry or its remnant.
 In condensed matter physics, such a question can be answered
  by changing the ground state properties of the system
  with temperature, pressure, doping etc as external controlling
 parameters \cite{AND}. If the  frequency  of 
 some collective mode has characteristic change near the 
 critical point, it provides us with  a clear signature
 of a direct link between the mode and the phase transition.
 The life is, unfortunately, not that easy in QCD
 from the experimental point of view.
 Nevertheless 
 the heavy nuclei provide us with finite  baryon-density 
 environment  which allows us to study the 
 partial restoration of chiral symmetry and associated 
 change of the meson properties.
 Heavy-ion collisions also provide us with not only
 high baryon density but also high temperature although
 the system stays in such environment only in a transient time. 

 Some years ago,  one of the present authors \cite{KUNI95} 
proposed several nuclear experiments including
 one using electro-magnetic probes for producing
  the $\sigma$ meson in nuclei
to see a clearer evidence of the $\sigma $ pole and to explore possible
 restoration of chiral symmetry in nuclear medium.
 When  a particular meson is  put in a nucleus,
 it may dissociate into complicated
 excitations to loose its identity in the  medium;
for example, $\sigma\leftrightarrow 2\pi$, 
  $p$-$h$, $\pi$+$p$-$h$,
 $\Delta$-$h$, $\pi$+$\Delta$-$h$,  $\cdots$.
Therefor it is not appropriate to talk about the 
 ``mass'' and ``width'' of the particle, but one
 should study the   spectral function which contains
 all those information.

\section{Chiral restoration and soft mode  at finite $T$ and density}

\newcommand{\con}{\la \bar{q}q\ra}
The basic observation on which the  whole discussions in this 
report are based is that the dynamical breaking of \chis is 
a phase transition of the QCD vacuum with an order 
parameter $\con $:
Then there may exist  collective excitations corresponding to 
the quantum fluctuations of the order parameter.
The phase (amplitude) fluctuation of the order parameter
 corresponds to  the pion (ur-$\sigma$).
 As we have mentioned before, they couple strongly and 
 mix through the process $\sigma \leftrightarrow 2 \pi$.
 
Now let us first show how  the magnitude of the 
condensate $\langle \bar{q}q\rangle $ 
decreases at finite  $T$ and/or density $n_{{B}}$ 
 in an analytic way.

\subsection{Chiral condensate at low $T$}

The condensate at $T \neq 0$ ($n_{_{B}} =0)$  can be written as
\beq
 \langle \bar{q}q \rangle   = \frac{1}{Z}
 {\rm Tr} \left[ \bar{q}q\   e^{- H_{\rm QCD}  /T} \right] 
  =  \frac{\partial f(T)}{\partial m_q}
   = - \frac{\partial P(T)}{\partial m_q},
\label{cond-T}
\eeq
where $Z$ is the QCD partition function,
$f$ is the  free energy density and $P$ is the 
the total pressure of the system
including vacuum pressure and the thermal pressure.
This is because the current quark masses enters 
$H_{\rm QCD}$ only in the  form $\sum_im_q\bar{q} q$;
hence this formula is valid for each  flavor ($q = u, d, s, c, b, t)$.

At high $T \gg \Lambda_{QCD}$, one can calculate $P(T)$ using
the high $T$ QCD perturbation theory and $ \langle \bar{q}q \rangle $
 is shown to be exactly zero for $m_q=0$. This is because 
 there is no vertex which flips chirality in the 
 QCD Lagrangian except for the mass term.
On the other hand, there is a dynamical breaking of
chiral symmetry at low $T$ and 
 $ \langle \bar{q}q \rangle $ does not necessarily
 vanish. For $T$ low enough, the system may be 
approximated by a dilute and weakly interacting gas of pions.
Its pressure can be  calculated using the chiral perturbation theory
 \cite{GL89}:
\beq
 Z =  {\rm Tr} \left[ e^{-H_{\rm QCD}/T} \right] =
\int [dU] \exp 
\left[ -\int_0^{1/T} d\tau \int d^3 x \ {\cal L}_{eff}(U) \right] ,
\eeq
where ${\cal L}_{eff}(U)$ is expanded
as  $ {\cal L}^{(2)} + {\cal L}^{(4)} + {\cal L}^{(6)} + \cdots $
in powers of $\partial /4\pi f_{\pi}$ and $m_{\pi}/4 \pi f_{\pi} $.
Therefore, 
 the quark mass dependence of $P(T)$ for $N_f=2$ 
 at low $T$ in (\ref{cond-T}) can be extracted from its $m_{\pi}$
dependence through the Gell-Mann-Oakes-Renner relation
\beq
m_{\pi}^2 = -2 \hat{m} \frac{\langle \bar{q} q \rangle_0 }{f_{\pi}^2}
   +  O(m_{u,d}^2) ,
\label{gor}
\eeq
where $\hat{m}=(m_u+ m_d)/2$, and 
   $\langle \bar{q} q \rangle_0$ and $f_{\pi}$
are the vacuum condensate and the pion decay constant 
in the chiral limit, respectively.

Writing $P(T) = P_{\rm pion}(T) + P_{\rm vac}$, and using
the above relation, one thus finds the quark-condensate 
in the chiral limit for $N_f=2$ as \cite{GL89}
\beq
\frac{ \langle \bar{q}q \rangle }
 {\langle \bar{q}q   \rangle_0 }
= 1 + \left. \frac{1}{f_{\pi}^2}
 \frac{\partial P_{\rm pion}(T)}{\partial m_{\pi}^2} 
  \right|_{m_{\pi} \rightarrow 0} 
= 1 - \frac{T^2}{8 f_{\pi}^2}
    - \frac{1}{6} \left( \frac{T^2}{8 f_{\pi}^2} \right)^2
    - \frac{16}{9} \left( \frac{T^2}{8 f_{\pi}^2} \right)^3
     \ln \left( \frac{\Lambda_q}{T} \right) + O(T^8) ,
\label{chicon}
\eeq
where $\Lambda_q (= 470 \pm 110 $ MeV) is a parameter
extracted from the experimental \pipi scattering
length in the isoscalar $D$-wave channel.
 The $T^2$ corrections in the above expansion originate
solely from the free pion-gas contribution.
Therefore, it can be  alternatively obtained by applying the 
soft pion theorem to the free pion gas.
 The negative contribution of the 
 $O(T^2)$ term is due to the fact that
 the pion has positive scalar-charge
  $\langle\pi(p)\vert\bar{q}q\vert\pi(p)\rangle $, 
 which can be seen from either by the soft-pion theorem or
 by the Feynman-Hellman theorem.
 On the other hand,
 the interaction among pions are relevant for the terms
 $O(T^{n \ge 4})$.

 The above result of the low $T$ expansion suggests that 
 the chiral condensate decreases uniformly as $T$ increases
 and the 
  symmetry restoration takes place somewhere around 100-200 MeV.
 However, the behavior of $\bar{q}q$ 
 near the critical point is out of reach of the low $T$ expansion,
 because the system is not any more the dilute gas of
 pions near the critical point.
 In such a region, non-perturbative methods such as the  
 the lattice QCD simulations are necessary to  
 make a reliable estimate of the chiral condensate.

\subsection{Chiral condensate at finite baryon density}

$\langle \bar{q}q \rangle$
at finite baryon density obeys  an exact formula in QCD 
 directly obtained from the Feynman-Hellman theorem 
\begin{eqnarray}
 \langle \Psi | \bar{q} q | \Psi \rangle 
 = \frac{\partial \langle \Psi |  {\cal H}_{QCD} | \Psi \rangle }
        {\partial m_q} ,
\label{FH-theorem}
\end{eqnarray}
which is valid for any eigenstates $| \Psi \rangle $ of the QCD Hamiltonian 
 $H_{QCD} (= \int d^3x {\cal H}_{QCD}$) 
with the normalization
 $\langle \Psi | \Psi \rangle =1$.
 For cold nuclear matter $| \Psi \rangle = | {\rm nm} \rangle $,
 the energy density is given as 
 \begin{eqnarray}
 \langle {\rm nm} |  {\cal H}_{QCD} | {\rm nm} \rangle
 = {\cal E}_{\rm vac} + n_B [ M_N + E_{\rm bin.} ],
\end{eqnarray}
 where ${\cal E}_{\rm vac}$ is the vacuum energy density,
 $M_N$ is the rest mass of the nucleon and
 ${\cal E}_{\rm bin.}$ is the binding energy of nuclear matter
 per particle. Applying this and the similar formula
 for the nucleon to (\ref{FH-theorem}), one arrives at   \cite{DL90}:
 \begin{eqnarray}
 \label{cond-rho}
{\langle \bar{u} u + \bar{d} d \rangle \over
 \langle \bar{u} u + \bar{d} d \rangle_{0} }
= 1 - {n_B \over f_{\pi}^2 m_{\pi}^2 }
 \left[ 
  \Sigma_{\pi N} + \hat{m} {d \over d\hat{m}} E_{\rm bin.}(n_B) 
\right] ,
\end{eqnarray}
where $\Sigma_{\pi N}= 45 \pm 10 $ MeV is the pion-nucleon sigma term.
The density-expansion of the r.h.s. in  the 
 lowest order
gives a reduction of almost 35 \% of $\langle \bar{q} q \rangle $
already at the nuclear matter density $n_0 = 0.17 $fm$^{-3}$:
The physical origin of this reduction is similar with the 
pion gas case discussed above; the nucleon  has
 the positive scalar charge $\la N | \bar{q}q | N \ra >0$, hence 
 the nucleon
 gas gives the positive contribution to the quark condensate.
The above estimate suggets that 
 heavy nuclei can be a good laboratory to
study the partial restoration of chiral symmetry
 in nuclear medium; 20 to 30\% reduction of the 
chiral condensate might  be large enough to manifest 
typical phenomena of 
partial restoration of chiral symmetry. We shall argue that 
this is actually the case.

\subsection{Hadronic correlations and symmetry restoration}

 The wisdom of many-body theory tells us that 
 the structure of the ground state is closely related with and 
 reflected in the excitation spectra of the system.
 In the case of chiral symmetry restoration,
 a model-independent statement on this connection goes as follows.
 Consider the  hadronic correlations
in a same chiral multiplet, such as $S$ (scalar)$-$$P^a$ 
(pseudo-scalar) 
and $V_{\mu}^a$ (vector)$-$$A_{\mu}^a$ (axial-vector). Then they
must be degenerate when $\langle \bar{q}q \rangle \rightarrow 0$,
namely,
\begin{eqnarray}
\langle S(x) S(y) \rangle 
 \rightarrow  \langle P^a(x) P^a(y) \rangle, 
\ \ \  \
\langle A^a_{\mu}(x) A^b_{\nu}(y) \rangle
  \rightarrow \langle V^a_{\mu}(x) V^b_{\nu}(y) \rangle.
\end{eqnarray}

\begin{figure}[t]
   \centering
   \epsfxsize=8.0cm
   \epsfbox{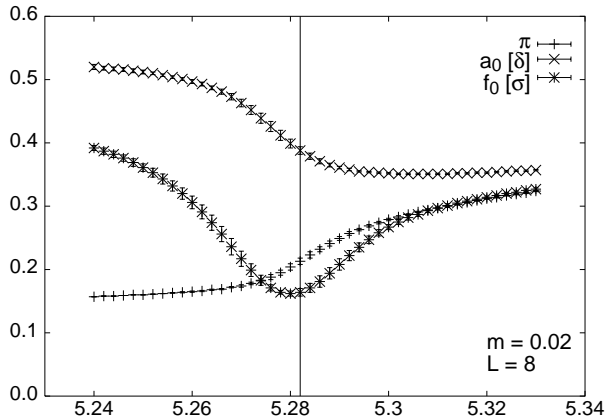}
   \vspace{0.5cm}
\caption{Thermal susceptibilities in three different channels
 ($\pi$, $\sigma$, and $\delta$) for two-flavor QCD on the
   $8^3 \times 4$ lattice with  $m_{u,d} \ a = 0.02$ {\protect \cite{KAR01}}.
 The vertical (horizontal) axis denotes $\sqrt{1/\chi_{_H}}$ 
(the lattice coupling  $\beta=6/g^2$).
}
\label{fig3}
\end{figure}

To see  such chiral degeneracy by lattice QCD simulations,
it is useful to define the thermal susceptibilities for hadronic
operators ${\cal O}$ defined in the Euclidean space,
\begin{equation}
\chi_{_H} = \int_0^{1/T}
 d\tau \int d^3x \  \langle {\cal O}^{\dagger}(\tau, \vec{x})
{\cal O} (0,\vec{0})\rangle .
\label{sus}
\end{equation}
Shown in Fig.\ref{fig3} is the lattice QCD simulation of 
$\sqrt{1/\chi_{_H}}$ with $N_f= 2$ dynamical quarks \cite{KAR01}.
One can see the degeneracy 
between the $\sigma$-channel ($I$=0, $J^P$=$0^+$)
and the $\pi$-channel ($I$=1, $J^P$=$0^-$)
above the critical point. 
Also,  $\sqrt{1/\chi_{_H}}$ for $\sigma$ is  smaller  
than that for $\delta$ at low $T$, which indicates 
that  $\sigma$ could be light. 
There remains  a splitting
between the $\delta$-channel ($I$=1, $J^P$=$0^+$)
and the $\pi$-channel even above the critical point, which reflects
the breaking of $U_A(1)$ symmetry in the 2-point
correlation function for $N_f=2$; in other words,
the $U_A(1)$ anomaly survives the chiral restoration.
In general, the effect of the  $U_A(1)$ symmetry breaking 
appears for $N (\ge N_f)$-point function for
$N_f$-quarks as shown in \cite{LH96}.

To make a direct connection 
 of the chiral restoration with
 the experimental observables, the Euclidean correlations 
 such as (\ref{sus}) are not useful enough;
one needs to extract the information of the 
 spectral distribution as we mentioned before.
Recently, a new approach has been proposed to
 extract hadronic  spectral functions (SPFs) from lattice
 QCD data by using the  maximum entropy method (MEM)   \cite{NAH99}.
 MEM  has been successfully applied for 
 similar problems in image reconstruction in crystallography and 
  astrophysics, and in 
  quantum Monte Carlo simulations in
 condensed matter physics \cite{LANL96}.
    
The Euclidean correlation function $D(\tau)$ of an 
operator ${\cal O}(\tau,\vec{x})$  and its spectral decomposition
at zero three-momentum read
\begin{eqnarray}
D(\tau > 0) = \int 
\langle {\cal O}^{\dagger}(\tau,\vec{x}){\cal O}(0,\vec{0})\rangle d^3
x\label{KA}  \equiv \int_{0}^{\infty} \!\! K(\tau,\omega) A(\omega ) d\omega,
\end{eqnarray}
where  $\omega$ is a real frequency, and
$A(\omega)$ is the spectral function 
(or sometimes called the {\em image}),
which is positive semi-definite. 
$K(\tau,\omega)$ is a known integral kernel
 (it reduces to Laplace kernel $e^{-\tau \omega}$ at $T$=0.)
 Monte Carlo simulation provides  $D(\tau_i)$ 
on  the discrete set of temporal points $0 \le \tau_i /a \le N_\tau$.
From this  data with statistical noise, we need to reconstruct the 
spectral function $A(\omega)$ with a continuous variable $\omega$. 
 This is a typical ill-posed problem, 
where the number of data is much smaller than
the number of degrees of freedom to be reconstructed.
MEM is a method to circumvent this difficulty
 through Bayesian  statistical inference of the most probable 
{\em image} together with its reliability.
There are three important aspects of MEM:
 (i) it does not require a priori
assumptions or parameterizations of SPFs, (ii) for given data, 
 a unique solution is obtained if it exists, and (iii)
the statistical significance of the solution can be 
quantitatively analyzed. 
 
\begin{figure}[tb]
   \centering
   \epsfxsize=12.0cm
   \epsfbox{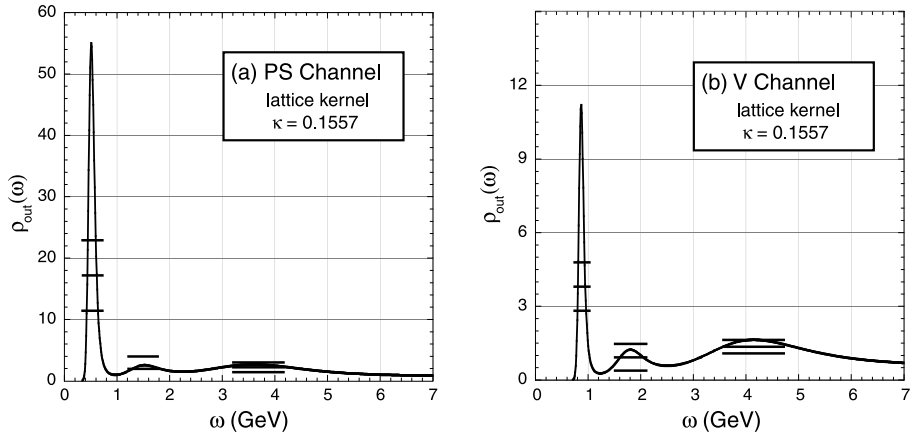}
   \vspace{0.5cm}
\caption{Spectral functions $\rho_{out}(\omega) 
 \equiv A(\omega) /\omega^2$
 obtained by MEM using  the quenched  lattice QCD 
 data. The lattice size is 20$^3$$\times$24 with $a=0.0847$ fm.
 12 data points in the temporal direction ($1 \le \tau_i /a \le 12$) are
  used for the MEM analysis. (a) is for the pion channel and (b) is 
  for the rho-meson channel. The figures are taken from
  {\protect \cite{NAH99}}.}
\label{fig4}
\end{figure}

In Fig.\ref{fig4}, shown is an example of the spectral function
in the pion and the rho meson channels at $T=0$ extracted from the 
quenched lattice QCD data \cite{NAH99}. 
Correct resonance peaks at low energies and 
the continuum structure at high energies are well reproduced.
The long-standing problem of in-medium spectral functions of
vector mesons
 ($\rho$, $\omega$, $\phi$, $J/\psi$, $\Upsilon$, $\cdots$, etc) 
and scalar/pseudo-scalar mesons ($\sigma$, $\pi$, $\cdots$, etc)
can be studied using MEM combined with finite $T$ lattice simulations.

\section{Scalar Correlation in the medium}

\subsection{Spectral enhancement  in hot matter}

The fluctuation of the order parameter   becomes large as
 the system approaches the critical point of a second order 
 or to the end point of the first order phase transition.
 This is easily seen  for the second order case by looking 
 at the Landau free-energy of a double-well type  
written in terms of the  order parameter 
$\sigma \sim \langle \bar{q}q \rangle $,
\begin{eqnarray}
V (\sigma) = - {a \over 2} \sigma^2 + {b \over 4} \sigma^4 ,
  \end{eqnarray}
where $a (\propto T_c - T)$
 changes  sign at the critical point, while $b$ remains
positive.
 The minimum value and the curvature at
 the minimum of the effective potential $\sigma_0$   read
\begin{eqnarray}
\sigma_0 = \sqrt {a \over b} , \ \ \ \ 
{1 \over 2} \left. {d^2 V(\sigma) \over d\sigma^2 }\right|_{\sigma = \sigma_0}
 = a \ \ ,
\end{eqnarray}
respectively.
 As $a$ becomes small, not only the order parameter
 $\sigma_0$ but also the curvature  decrease.
 This is nothing but the softening of the oscillational mode 
of the order parameter associated with the second order phase transition.

\subsubsection{Soft mode in the Wigner phase}

\begin{figure}[tbp]
   \centering
   \epsfxsize=7.0cm
   \epsfbox{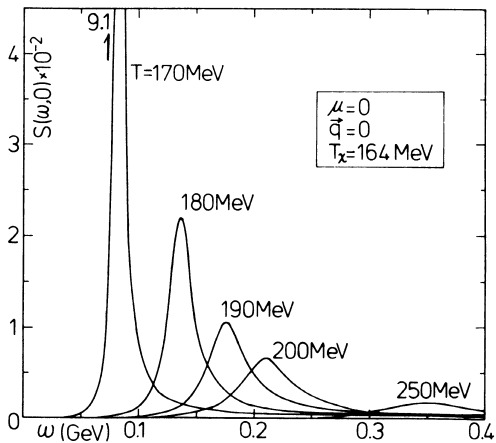}
 \vspace{0.5cm}
 \caption{The spectral function of degenerate ``para-pion''
  and ``para-sigma" at finite $T$ in the Wigner phase
 calculated by the 2-flavor 
 Nambu-Jona-Lasinio model in the chiral limit.
 Critical temperature of the chiral transition is 164 MeV 
 and the spectral enhancement is seen as $T$ approaches
 to the critical point. Taken from {\protect \cite{HK85}}.}
    \label{fig5}
\end{figure}

Historically, the dynamical fluctuation of the 
 order parameter near the chiral critical point
 was first considered in the Wigner phase above $T_c$
 by using the Nambu-Jona-Lasinio model \cite{HK84,HK85,ptp85}.
 It was shown that such collective oscillation becomes soft and
  narrow as the system approaches to the critical point.
  Since this is the collective mode in the chirally symmetric phase,
  it may be called ``para-pion'' or ``para-sigma''
 in analogy to the para-magnon
  in the electron gas and 
liquid $^3$He \cite{doniach} 
and para-superconductivity in low dimensional
  superconductors\cite{paracon}.
In nuclear physics, the soft mode for the superconductivity 
is called the pairing vibration,  and the $2^{+}$ phonon
is the soft mode for the quadrupole deformation of nuclei.\cite{bohr}
The soft mode for the pion condensation \cite{TAMA93}
should be the longitudinal spin-dependent isovector modes
with finite momentum\cite{sigmatau,TAMA93}.
  In Fig.\ref{fig5}, shown is the
   spectral function of the para-pion defined by 
 \beq
 S^{ab} (\omega, {\bf p} ;T) \propto {\rm Im} \int d^4x \ e^{ipx}\  \langle
 {\rm R} \ \bar{q} \Gamma^a q(x) \bar{q} \Gamma^b q(0) \rangle  ,
\eeq
 where $\Gamma^a = (1, i \gamma_5 \vec{\tau} )$ and ${\rm R}$
 denotes the retarded product\cite{HK85}.

\subsubsection{Soft mode in the Nambu-Goldstone phase}

In the chirally broken phase just below $T_c$,
there arises two types of soft modes, namely
the phase and amplitude fluctuations of the chiral order parameter 
 $\langle \bar{q}q \rangle $\cite{HK85,ptp85}.
Because of the Nambu-Goldstone theorem, the pion
stays massless in the chiral limit below $T_c$, while 
one can expect  sizable   softening (the red-shift) of  $\sigma$.
In the real world, the situation is not that simple,
 since  $\sigma$ has
 a large width from the decay $\sigma \rightarrow 2 \pi$.
Nevertheless,   $\sigma$ may 
appear as a narrow resonance near $T_c$ because of the
phase space suppression of the above decay channel, 
although it is at best 
a  broad resonance in the free space.
This idea has been put forward in \cite{HK87} for the first time 
and various  phenomenological applications 
in relation  to the relativistic heavy ion collisions
have been discussed later \cite{WEL92}.
 
Since the character of $\sigma$ changes from a very broad
 resonance to a sharp resonance, it is not at all
  enough to just talk about the mass and width but is
 necessary to study   the 
 spectral function.
 By using the linear $\sigma$-model, the spectral functions
  in $\pi$ and $\sigma$ channels have been studied at finite $T$
 and it was found that there is a significant spectral enhancement
 just above the 2$m_{\pi}$ threshold as $T$ increases \cite{CH98}. 

To describe the general
 features of this spectral enhancement, let us 
 consider the propagator 
 of the scalar-isoscalar $\sigma$-meson at rest in the medium:
\begin{eqnarray}
D^{-1}_{\sigma} (\omega)= \omega^2 - m_{\sigma}^2 - 
\Sigma_{\sigma}(\omega;T),
\end{eqnarray}
where $m_{\sigma}$ is the mass of the $\sigma$ in the tree-level, and
$\Sigma_{\sigma}(\omega; T)$ is 
the loop corrections
in the vacuum as well as in the hot medium.
The corresponding spectral function is given by 
$\rho_{\sigma}(\omega) = - {1 \over \pi } {\rm Im} D_{\sigma}(\omega).$
Near the 2$m_{\pi}$ threshold, the imaginary part 
 in the one-loop order	 reads 
\begin{eqnarray}
{\rm Im} \Sigma_{\sigma} \propto \theta(\omega - 2 m_{\pi}) \ 
	 \sqrt{1 - {4m_{\pi}^2 \over \omega^2}} .
\end{eqnarray}

When chiral symmetry is being restored,
$m_{\sigma}^*$ (``effective mass'' of $\sigma$ 
defined as a zero of the real part of the propagator
 ${\rm Re}D_{\sigma}^{-1}(\omega = m_{\sigma}^*)=0$)
 approaches to $ m_{\pi}$.  Therefore,
 there exists a temperature at which 
 ${\rm Re} D_{\sigma}^{-1}(\omega = 2m_{\pi})$
 vanishes even before the complete $\sigma$-$\pi$
 degeneracy is realized; 
$ {\rm Re} D_{\sigma}^{-1} (\omega = 2 m_{\pi}) =
 [\omega^2 - m_{ \sigma}^2 -
 {\rm Re} \Sigma_{\sigma} ]_{\omega = 2 m_{\pi}} = 0.$
At this point, 
the spectral function can be  solely represented by the
 imaginary part of the self-energy;
\begin{eqnarray}
\rho_{\sigma} (\omega \simeq  2 m_{\pi}) 
 =  - {1 \over \pi \ {\rm Im}\Sigma_{\sigma} }
 \propto {\theta(\omega - 2 m_{\pi}) 
 \over \sqrt{1-{4m_{\pi}^2 \over \omega^2}}},
\end{eqnarray}
which shows an enhancement of the spectral function at the $2\pi$
threshold. 
One should note that this enhancement is 
 due  to the  partial restoration of chiral symmetry and hence 
generic \cite{CH98}.

To make the argument more quantitative,
let us evaluate $\rho_{\sigma}(\omega)$ in 
 the O(4) linear $\sigma$-model:
\begin{eqnarray}
\label{model-l}
{\cal L}_M  =  {1 \over 4} {\rm tr} [\partial M \partial M^{\dagger}
 - \mu^2 M M^{\dagger} 
  - {2 \lambda \over 4! } (M M^{\dagger})^2   - h (M+M^{\dagger}) ],
\end{eqnarray}
where tr is for the flavor index and  
$M = \sigma + i \vec{\tau}\cdot \vec{\pi}$.
Although the model has only a limited number of parameters
 and is not a precise low energy representation of QCD
at zero $T$, 
it can  describe the
pion dynamics qualitatively well up to 1GeV  \cite{CH73}.
It also provides a good description of the 
static critical phenomena of QCD  owing to the universality 
 argument \cite{PW84,WIL92}.
 The coupling constants $\mu^2, \lambda$ and $h$ have
been determined in the
vacuum to reproduce $f_{\pi}=93$ MeV, $m_{\pi}=140$ MeV as well as
the s-wave $\pi$-$\pi$ scattering phase shift in the one-loop order:
 one of such parameter sets is
 $-\mu^2 = (284 {\rm MeV})^2$, $\lambda/4\pi = 5.81$ and
 $h^{1/3} = 123 {\rm MeV}$ \cite{CH98}. This will be used through out
 this report.
 
\begin{figure}[t]
    \centering
   \epsfxsize=12.0cm
   \epsfbox{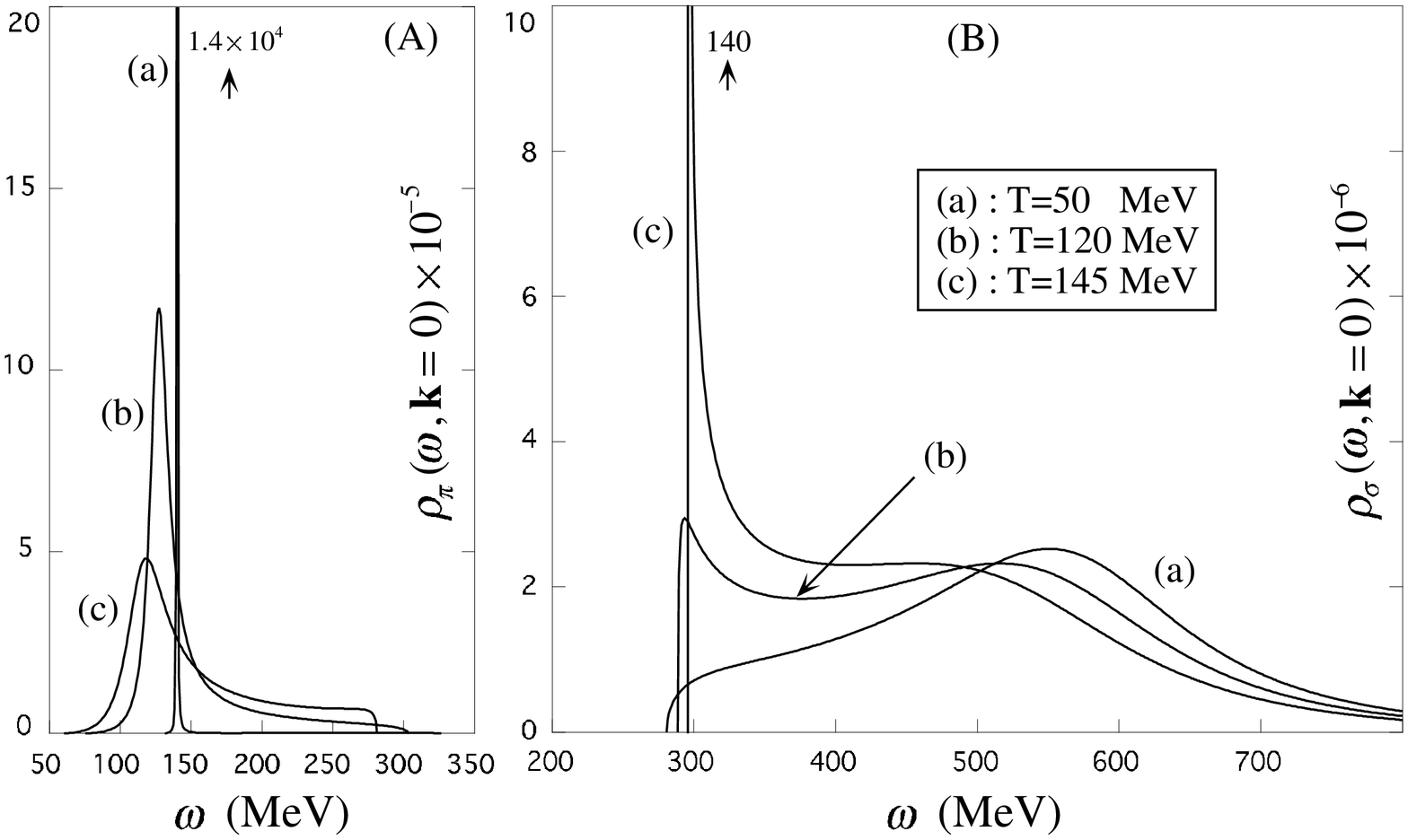}
 \vspace{0.5cm}
 \caption{ Spectral functions in the $\pi$ channel (A) and in the
 $\sigma$ channel (B) for $T$=50, 120,  and 145 MeV with the
 parameter set shown in the text  {\protect \cite{CH98}}.
 $\rho_{\sigma}(\omega)$ in (B) shows only a broad bump at low $T$ (a), while
 the spectral concentration is developed as $T$ increases,
  (a)$\rightarrow$(b)$\rightarrow$(c).}
  \label{fig6}
\end{figure}

In Fig.\ref{fig6}, shown are the 
spectral functions $\rho_{\pi (\sigma)}$ 
in the $\pi$ ($\sigma$)-channel
at finite $T$ calculated in the $O(4)$ linear $\sigma$ model:
Two characteristic features are 
the broadening of the pion peak (Fig.3(A)) and the
spectral concentration at  the 2$m_{\pi}$ threshold 
 ($\omega \simeq  2m_{\pi}$)  in the $\sigma$-channel (Fig.3(B)).
The latter may be measured by the $2\gamma$
spectrum from the hot plasma created in the 
relativistic heavy ion collisions \cite{CH98,VOL98}.

\subsection{Spectral enhancement  in nuclear matter}
 
\begin{figure}[t]
    \centering
 \begin{minipage}{0.49\textwidth}
   \epsfxsize=8.0cm
   \epsfbox{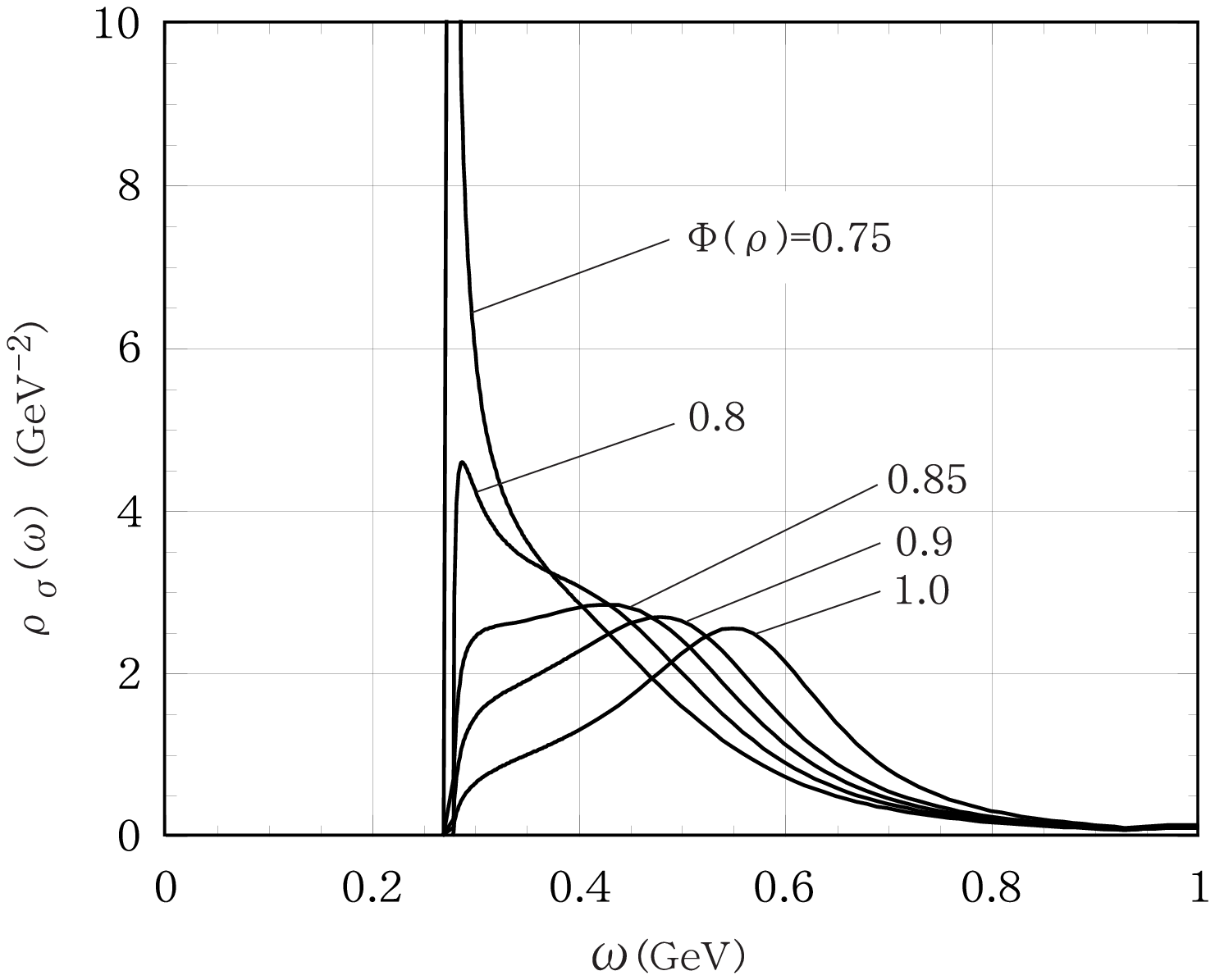}
 \end{minipage}
 \begin{minipage}{0.49\textwidth}
 \hspace{1cm}
   \epsfxsize=6.5cm
   \epsfbox{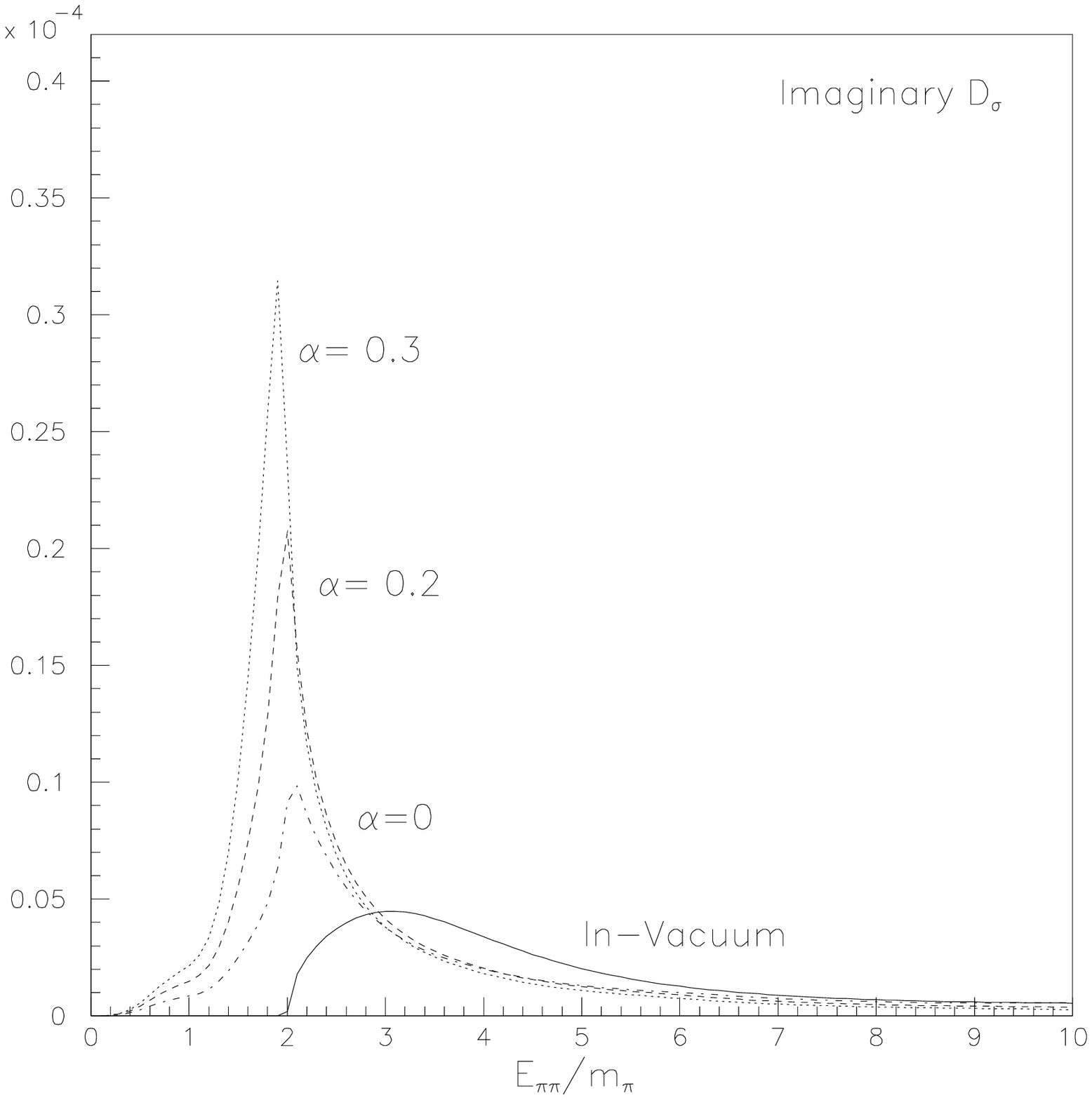}
 \end{minipage}
 \vspace{0.5cm}
\caption{Left panel : The spectral function in the 
 $\sigma$ channel for several values of 
 $\Phi = \langle \sigma \rangle / \langle \sigma_0 \rangle $
 with the same parameter set of the 
 linear $\sigma$ model  as Fig.(\ref{fig6}). 
 This figure is taken from {\protect \cite{HKS99}}.
 Right panel : The phenomenological
 spectral function in the 
 $\sigma$ channel with the p-wave effects such as
 $p$-$h$ and $\Delta$-$h$ excitations.
 The solid line corresponds to the vacuum case and the dashed lines
 correspond to the in-medium case at $\rho = \rho_0$
 with several values of 
 $\alpha (\equiv 1 - m_{\sigma}^*(\rho=\rho_0)/m_{\sigma}^*(\rho=0))$, 
 {\protect \cite{SCHUCK00}}. }
\label{fig7}
\end{figure}

The idea of the spectral enhancement near the 2$m_{\pi}$ threshold
can be also applied to the finite density system \cite{HKS99}.
A surprise was that such enhancement can be
 seen even near the nuclear matter density if
 the chiral condensate decreases by 30 \% or so.
 To see this explicitly, 
 let us   consider again    the  O(4) linear $\sigma$ model
to  demonstrate the essential idea of the near-threshold enhancement.
The Lagrangian adopted in \cite{HKS99} is the standard Gell-Mann-Levy
model with a minor modification
\beq
\label{model-ll}
{\cal L}_{MN} = {\cal L}_{M} +
 \bar{\psi} ( i \delslash - g M_5 ) \psi  + \cdot \cdot \cdot  ,
\eeq
where  $M = \sigma + i \vec{\tau}\cdot \vec{\pi}$,
$M_5 = \sigma + i \gamma_5 \vec{\tau}\cdot \vec{\pi}$,
and  $\psi$ is the nucleon field.
In the mean-field level, medium effect originating from the nucleons
can be incorporated by making the following replacement;
\beq
\la \sigma \ra_0 \rightarrow \la \sigma \ra
 =  \ \Phi (\rho) \la \sigma \ra_0  ,
\label{replace}
\eeq
where $\la \cdot \ra_0$ ($\la \cdot \ra$) denotes the
vacuum (nuclear matter) expectation value, and
$\rho$ ($\rho_0$) is the baryon density (nuclear matter
 saturation density).\footnote{$\rho$ in the following
 corresponds to $n_B$ in previous sections.}
$\Phi(\rho)$ may be parameterized as
 $\Phi(\rho) = 1 - C \rho / \rho_0 $ with $C = 0.1 -0.3$
 at low density.
The resultant spectral function in the $\sigma$ channel
is shown in Fig.\ref{fig7} (left panel).
 The spectral enhancement just above the 2$m_{\pi}$ threshold
 similar to the case at finite $T$ (Fig.\ref{fig6}(B))
 can be seen when $\langle \sigma \rangle$ changes by 25 \% from its
 vacuum value. 
 This mean-field treatment was later improved by including
 the $p$-$h$ and $\Delta$-$h$ contributions to the pion 
 propagator \cite{SCHUCK00}. In this case, the spectral strength
 spreads into the energy region even below 2$m_{\pi}$, but
 the qualitative feature of the enhancement at 2$m_{\pi}$
 still remains as shown  in Fig.\ref{fig7} (right panel)
 where  $\alpha \equiv 1 -
m_{\sigma}^*(\rho=\rho_0)/m_{\sigma}^*(\rho=0)$.

One can also study the in-medium
$\pi$-$\pi$ amplitude using the same model \cite{JHK01}.
A simple unitarized $\pi$-$\pi$ amplitude
in the $I$=$J$=0 channel reads
\begin{equation}
   \label{newT}
 T(s) = {1  \over  T_{\rm tree}^{-1}(s) - i \Theta(s)} ,
\end{equation}
where $T_{\rm tree}(s)$ is the in-medium tree-level amplitude
 with the mean-field replacement (\ref{replace}), and 
 $\Theta(s)$ is the phase space factor defined as
\footnote{Here we have changed the definition of 
 $T(s)$ by factor 2 from that in \cite{JHK01}.}
\beq
\label{thetas}
\Theta(s) = \theta(s - 4 m_{\pi}^2 ) \frac{1}{16 \pi}
 \sqrt{1-\frac{4 m_{\pi}^2}{s}} .
\eeq

\begin{figure}[b]
   \centering
 \begin{minipage}{0.45\textwidth}
   \epsfxsize=8cm
   \epsfbox{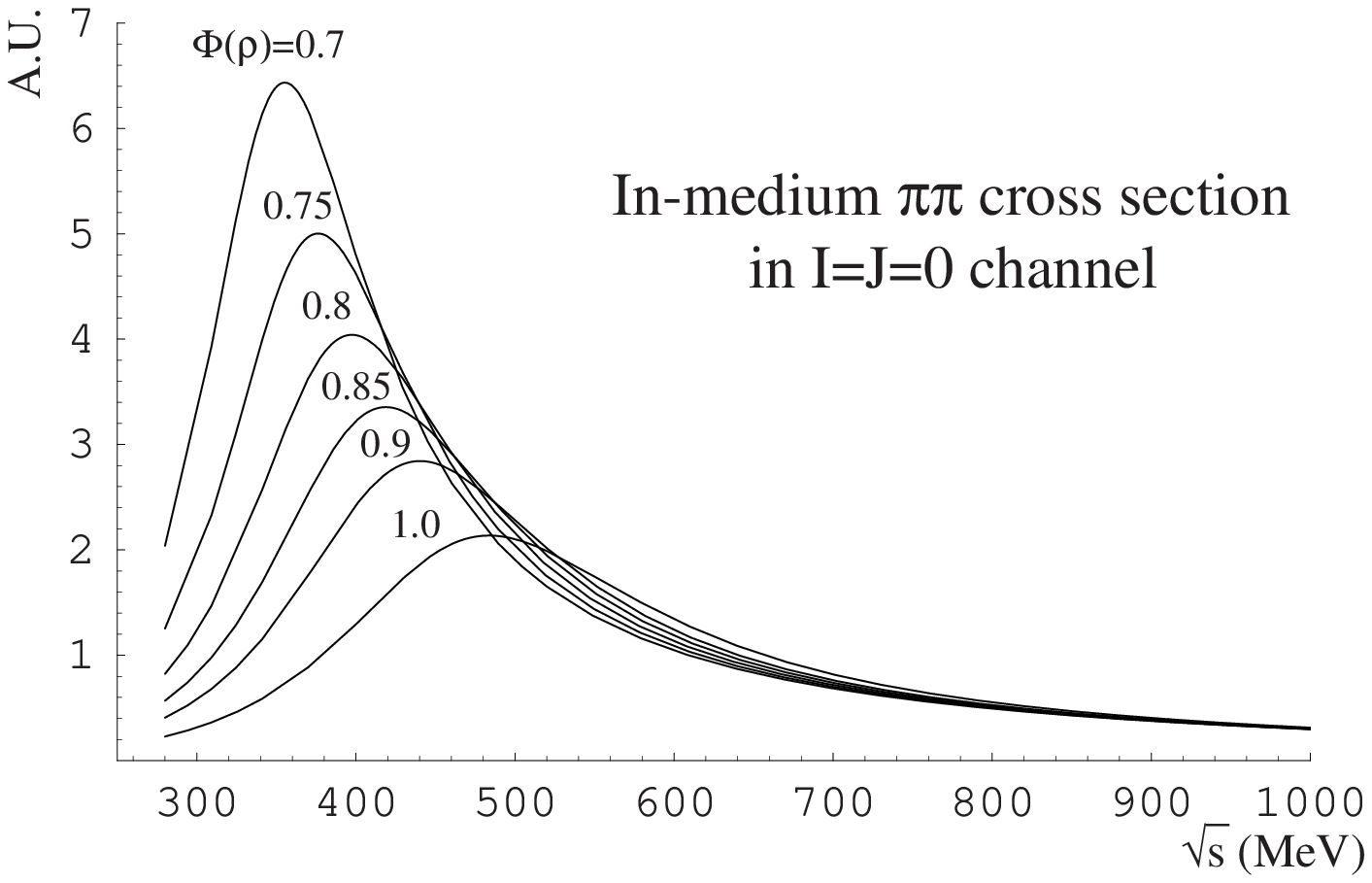}
 \end{minipage}
 \hspace{0.3cm}
 \begin{minipage}{0.45\textwidth}
   \epsfxsize=7.5cm
   \epsfbox{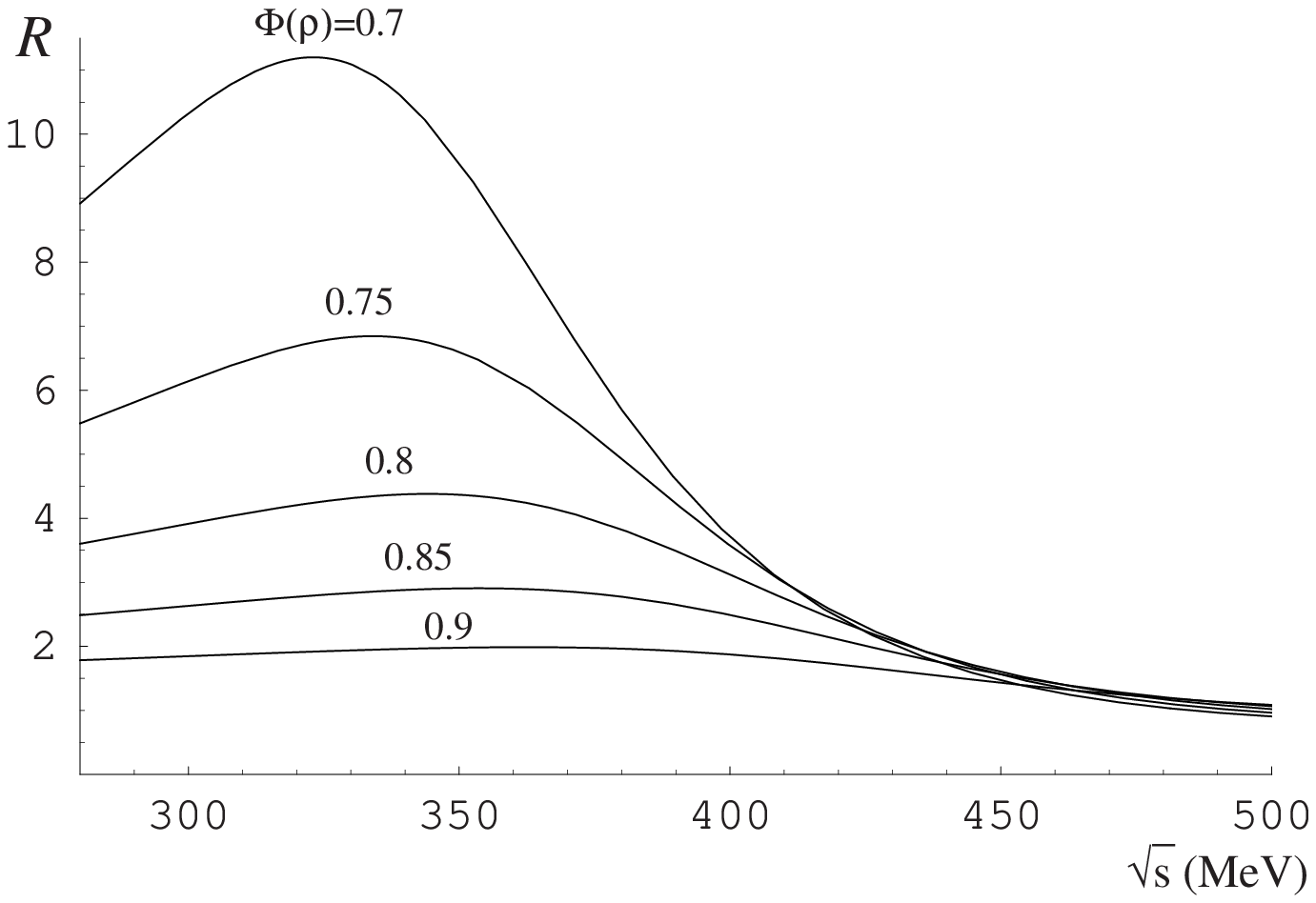}
 \end{minipage}
 \vspace{0.5cm}
   \caption{Left panel: In-medium $\pi$-$\pi$ cross section calculated 
 in the linear $\sigma$ model in the mean-field approximation
 {\protect \cite{JHK01}}.
 The same parameter set with Fig.(\ref{fig6},\ref{fig7}) 
 are taken. Right panel : 
 In-medium enhancement of the cross section relative to 
 its vacuum value in the range $2 m_{\pi} \leq \sqrt{s} \leq 500$ MeV
  {\protect \cite{JHK01}}. }
    \label{fig8}
\end{figure}

In Fig.\ref{fig8} (left panel),
 the in-medium \pipi cross section $\sigma_{\pi \pi} (s; \rho)$
 is shown in the arbitrary unit, in which the same parameter sets
 as in Fig.\ref{fig6} and Fig.\ref{fig7} are used.
 Again a large enhancement of the cross section 
 near the threshold is seen as the baryon density increases.
 In Fig.\ref{fig8} (right panel), shown is the 
 $\sigma_{\pi \pi} (s; \rho)$
relative to its vacuum value defined as 
\beq
\label{R-ratio}
R={ \sigma_{\pi\pi}(s;\rho)
  \over \sigma_{\pi\pi}(s;\rho=0)}.
\eeq

A  large enhancement near 2$m_{\pi}$ threshold
  in Fig.\ref{fig8} can be understood as follows:
 As the partial restoration of chiral symmetry takes place
 and  $\la \sigma \ra$ decreases,
 $m_{\sigma}^*$   approaches
 2$m_{\pi}$, which implies that $T_{\rm tree}^{-1}(s \simeq 2 m_{\pi})$
 tends to be  suppressed. Hence the $s$-dependence of
 the full inverse amplitude $T^{-1}(s)$ just above the 
2$\pi$ threshold is governed 
 by the imaginary part $\Theta(s  ) /2$, which causes the
  near-threshold enhancement of $T(s)$.

 At this point, a natural question to ask is
 whether the near-threshold enhancement obtained in the O(4)
 linear $\sigma$
 model arises also in the non-linear models or in the models which do
not have $\sigma$ explicitly \cite{JHK01}. 
 Furthermore, if it is the case,
 what kind of vertices in the non-linear chiral Lagrangian
 are responsible for the enhancement?
 To study these problems,
it is best to start with
 the standard polar parameterization of the chiral 
field\cite{JHK01},
 $M = \sigma + i \vec{\tau} \cdot \vec{\pi}
  = (\la \sigma \ra + S) U $ with $U = \exp (i \vec{\tau}
 \cdot \vec{\phi} /f^{*}_{\pi})$. Here,
$\la \sigma \ra$ is the chiral  condensate in nuclear matter
 as before, while
 $f^{*}_{\pi}$ is an ``in-medium pion decay constant''.
 The nucleon field is also transformed from $\psi$ to the
 chiral invariant field $N$.

Now let us take a heavy-scalar ($S$) limit and 
heavy-baryon ($N$) limit simultaneously.
 These limits can be achieved by taking 
 $\lambda, g \rightarrow \infty$ with
 $g/\lambda$ and $\la \sigma \ra_0 = f_{\pi}$ being fixed.
 In this limit,
 the heavy scalar field  $S$ may be integrated out
 and the following effective Lagrangian is obtained:
\beq
\label{model-nl2}
{\cal L}_{\rm eff}  = 
 \left(
 {f_{\pi}^2 \over 4}
 - {g f_{\pi}  \over 2 m_{\sigma}^2}\bar{N}N
 \right)
\left(
{\rm Tr} [\partial U \partial U^{\dagger}]
 - {h \over f_{\pi}} \  {\rm Tr}[U^{\dagger}+U] \right)
 \ + \  {\cal L}_{\pi N}^{(1)} + \cdot \cdot \cdot \ \ ,
\eeq
where all the constants take their vacuum values:
 $f_{\pi} = \la \sigma \ra_0 $,
 $m_{\sigma}^2 = \lambda \la \sigma \ra_0^2/3 + m_{\pi}^2$,
 and  $m_N = g \la \sigma \ra_0$.
 Note that $g f_{\pi} / 2  m_{\sigma}^2$
 in front of $\bar{N}N$ approaches to a finite value
$3 g/2\lambda f_{\pi}$ in the heavy limit, 
thus it cannot be neglected.
In (\ref{model-nl2}), 
${\cal L}_{\pi N}^{(1)}$ is the standard $p$-wave $\pi$-$N$ coupling
 and  $\cdot \cdot \cdot$  denotes
  other higher dimensional operators which are not relevant for the
   present discussion.

\begin{figure}[bth]
   \centering
   \epsfxsize=3.0cm
   \epsfbox{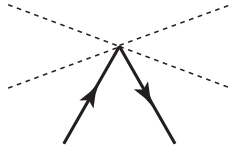}
 \vspace{0.5cm}
    \caption{An example of the new 4$\pi$-$N$-$N$ vertex generated
 by ${\cal L}_{\rm new}$.
 The solid line with arrow and the dashed line represent
 the nucleon and the pion, respectively.}
    \label{fig9}
\end{figure}

In the uniform nuclear matter, $\bar{N}N$ in
 eq.(\ref{model-nl2}) may be replaced by $\rho$.
This leads to a reduction of  the vacuum condensate;
$f_{\pi} = \la \sigma \ra_0  \rightarrow
  \la \sigma \ra =
 \la \sigma \ra_0 (1- g  \rho/f_{\pi} m_{\sigma}^2)
 = f_{\pi}^*$.
Then the proper normalization of the
pion field  in nuclear matter should be
 $\phi' = (\phi /f_{\pi}) \cdot f_{\pi}^* $.
Thus, the origin of the
 near-threshold enhancement in the heavy  limit
 can be ascribed to the following  new vertex:
\beq
\label{new-vertex}
{\cal L}_{\rm new} = - {3g \over 2 \lambda f_{\pi}} \
\bar{N}N {\rm Tr} [\partial U \partial U^{\dagger}].
\eeq
Because this vertex is proportional to
 the scalar-isoscalar density of the nucleon,
  it affects not only the pion propagator  but also
 the interaction among pions in nuclear matter.
 In Fig.\ref{fig9},  4$\pi$-$N$-$N$ vertex generated by ${\cal L}_{\rm new}$
 is shown as an example.
 Note that the vertex in Fig.\ref{fig9} 
acts to enhance the $\pi\pi$ attraction in
 the $I$=$J$=0 channel in nuclear matter. Despite its important role,
 this vertex has not been considered so far  in the calculations
of  the $\pi\pi$ scattering amplitudes in nuclear matter in the
non-linear approaches.

Here it should be pointed out that
 the vertex eq.(\ref{new-vertex})
 has been known to be one of the next-to-leading order terms
 in the non-linear chiral Lagrangian in the
heavy-baryon formalism  \cite{GSS88}.
In fact,  
  Thorsson and Wirzba \cite{TW95} have derived a general in-medium
 chiral Lagrangian by taking the mean-field
 approximation for the nucleon field;
\beq
\label{mean-f}
\la {\cal L} \ra
  =  ( {f_{\pi}^2 \over 4} + {c_3 \over 2} \rho)\  {\rm Tr}
 [\partial U \partial U^{\dagger}] 
   \ \   + \ \  ( {c_2 \over 2} - {g_A^2 \over 16 m_N})\ \rho \  {\rm Tr}
 [\partial_0 U \partial_0 U^{\dagger}] \\ \nonumber
\ \ \ \ \ \ \ \ \ \ \ \ \ \ \ \ \ \ \ \ \ \ \ \ \ \ \ \ \ 
 +  \ \
  ({ f_{\pi}^2 \over 4} + {c_1 \over 2} \rho)
 \ {\rm Tr} (U^{\dagger}  \chi + \chi^{\dagger} U).
\eeq
Note that ${\cal L}_{\pi N}^{(1)}$
 disappears in (\ref{mean-f}) in the {\it mean-field level}
 because of the isospin symmetry in nuclear matter, although it
 may contribute beyond the mean-field approximation as pointed
 out in \cite{MOW01}.
As is shown in \cite{GSS88}, the coefficients $c_{1,2,3}$ are related
to the $\pi N$ sigma term, the iso-spin even
 $S$-wave $\pi N$ scattering length, and
 the nucleon axial polarizability, respectively.
 We just quote here the numbers although they  have large potential
 uncertainties:
 $c_1 = -0.87 {\rm GeV}^{-1}$,  $c_2 = 3.34 {\rm GeV}^{-1}$,
 and  $c_3 = -5.25 {\rm GeV}^{-1}$.

By comparing eq.(\ref{model-nl2}) with
eq.(\ref{mean-f}), we observe that
the new vertex of the form
 $\bar{N}N {\rm Tr} [\partial U \partial U^{\dagger}]$
 appear in both cases with the same sign.
 In  eq.(\ref{model-nl2}), we have only the Lorentz invariant terms
 because we have integrated out only the scalar meson $S$.
 On the other hand,
 eq.(\ref{mean-f}) contains more general $O(3)$ invariant
 terms because it potentially contains the effect of
 heavy excited baryons \cite{GSS88}.
  It is of
  of great importance to make
 an extensive analysis of the $\pi \pi$ interaction in nuclear matter
 with  the new 4$\pi$-$N$-$N$ vertex.
 
\subsection{Behavior of the $\sigma$ pole in 2nd Riemann sheet}

 We  emphasize here 
 the relation of the near-threshold enhancement and the
 softening of the  fluctuating mode in the $\sigma$ channel in
 nuclear matter. We shall show that
such a fluctuating mode can be characterized by a
 complex pole of the unitarized amplitude $T(s)$  in eq.(\ref{newT}).

To see this in a simplest way, 
let us first take the heavy scalar limit 
($m_{\sigma} \rightarrow \infty$)  and 
  the chiral limit ($m_{\pi} \rightarrow 0$). 
In this case, 
the \pipi correlation is generated only by the 
standard lowest order chiral amplitude in eq.(\ref{newT})
 with $T_{\rm tree}(s) = s/f_{\pi}^2$.
 Then one readily finds  a pole of $T(s)$ 
 in the lower half plane as \cite{JHK01}
\beq
 \sqrt{s_{\rm pole}} & = & \sqrt{8\pi} \la \sigma \ra \ (1-i),
 \nonumber \\
 & = & (466 -466 i) \frac{\la \sigma \ra}{\la \sigma \ra_0} \ {\rm MeV}.
\label{simple-r}
\eeq
This complex pole represents a broad damping-mode dynamically
 generated by the $\pi\pi$ multiple scattering 
in the $I$=$J$=0 channel.
Since $\la \sigma \ra$ decreases in nuclear matter, the pole
 moves toward the origin in the complex
 $\sqrt{s}$-plane \cite{Oller} and thus causes an enhancement in the
 $\pi\pi$ cross section as shown in Fig.\ref{fig8}.
The existence of such  softening in the complex plane 
and its connection with 
 the near-threshold enhancement of the spectral 
function  was  first shown  in \cite{HK84,ptp85} using
the Nambu-Jona-Lasinio model. 

We shall now see that these qualitative features,
i.e., the existence and the softening of a complex pole do not 
change in the more realistic case with the real part of the 
rescattering amplitude included properly.
 We start again
with  the non-linear model with $m_{\sigma}\rightarrow \infty$.
The  $T$ matrix resumed in the inverse amplitude method\cite{IAM} reads
\beq
T(s)=T_2/(1+T_2g(s)),
\eeq
where $T_2$ is the lowest order chiral amplitude
 ($=s/f_{\pi}^2$ in the chiral limit) and $g(s)$ is the pion loop integral;
\beq
g(s)=i\int\frac{d^4 q}{(2\pi)^4}\frac{1}{q_{+}^2-m_{\pi}^2+i\eps}
\frac{1}{q_{-}^2-m_{\pi}^2+i\eps},
\eeq
with $q_{\pm}=q\pm p/2$ and $s=p^2$.
We notice that
${\rm Im}g(s)=-\Theta (s)$, which is given in (\ref{thetas}).
The real part of $g(s)$ is logarithmically divergent, so
 one needs a renormalization or
 use the one-subtracted dispersion relation, which 
introduce an arbitrary constant.
Actually, this program can be nicely formulated in the 
$N/D$ method with the dimensional regularization\cite{OOP98}.
However, as is also shown in \cite{OOP98}, the resultant 
formula can be obtained by a simple cut-off method 
provided that  the cut-off value is properly chosen.
The result in the chiral limit is given by \cite{OOP98,Oller}
\beq
g(z) =\frac{1}{16\pi^2}\{\tilde{a}^{SL} -{\rm Log}(- \mu^2/z)\},
\eeq
for $z$ in the first sheet, and 
\beq
g(z) =\frac{1}{16\pi^2}\{\tilde{a}^{SL} -{\rm Log}(- \mu^2/z)+2\pi i\}, 
\eeq
for $z$ in the second sheet  ({\rm Im} $z<0$).
Here ${\rm Log}z$ is the principal value of the logarithm;
it has a cut along the real negative axis and 
$-\pi<{\rm Arg}[{\rm Log}z]<\pi$. 
The constant $\tilde{a}^{SL}$
 is a cut-off dependent constant and is determined so that the
empirical phase shift in the scalar channel is reproduced
\cite{OOR}; $\tilde{a}^{SL}\simeq -.75\pm .2$ with $\mu=760$ MeV.
The pole position is given by the following dispersion equation
\beq
T_2^{-1}(z)=-g(z).
\eeq
 The result is shown in Fig.5: When $f_{\pi}=93$ MeV, 
the pole exists at $\sqrt{s}=461  -341 i$ MeV.
As $f_{\pi}$=$\langle \sigma \rangle$ is decreased,
 the pole  moves toward the origin.
It means that the mass and the width of the $\sigma$  mode decreases
as the chiral symmetry is restored; thus we confirm that the $\sigma$ is the
soft mode of the chiral transition.

\begin{figure}[bth]
   \centering
   \epsfxsize=8.0cm
   \epsfbox{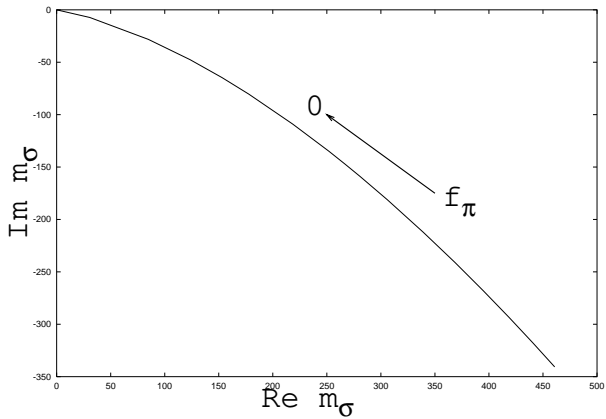}
 \vspace{0.5cm}
\caption{The shift of the pole position of the $T$ matrix 
in the chiral limit as $f_{\pi}$=$\langle \sigma \rangle$ 
 decreases toward $0$. }
\label{fig10}
\end{figure}

\section{Possible experimental signature}

\subsection{Softening in the scalar channel}

The CHAOS collaboration 
  measured the 
$\pi^{+}\pi^{\pm}$
invariant mass distribution $M^A_{\pi^{+}\pi^{\pm}}$ in the
 reaction $A(\pi^+, \pi^{+}\pi^{\pm})A'$ with 
 the incident pion momentum $p_{\pi^+} = 399$ MeV
 and  the targets  $A$=$^2$H, $^{12}$C, $^{40}$Ca, $^{208}$Pb
 \cite{CHAOS00}. They observed that
the   yield for  $M^A_{\pi^{+}\pi^{-}}$ 
 near the 2$m_{\pi}$ threshold is close to zero 
for $A=2$, but increases  with increasing $A$. They
identified that the $\pi^{+}\pi^{-}$ pairs in this range of
 $M^A_{\pi^{+}\pi^{-}}$ is in the $I$=$J$=0 state.

Although this experiment was originally motivated by the 
 possible strong $\pi$-$\pi$ correlation in 
 nuclei due to many-body effects \cite{GSI88},
 state of the art calculations based on the 
conventional many-body theoretical approach without
 incorporating the effect of partial restoration of 
 chiral symmetry do not
reproduce the sufficient enhancement in the near-threshold
 region in the $I$=$J$=0 channel \cite{RAPP99}.
 However, once the effect of partial chiral restoration  in nuclei
 suggested in \cite{HKS99} is incorporated together with
 the conventional approach, 
the near threshold enhancement
 enough to explain the experimental data
 may be obtained \cite{SCHUCK00}. So far such an analysis is still
 in the qualitative level. 
Quantitative calculations based on both the 
linear and non-linear approach discussed in the previous
 section are called for.
 
\begin{figure}[tbp]
   \centering
   \epsfxsize=7.0cm
   \epsfbox{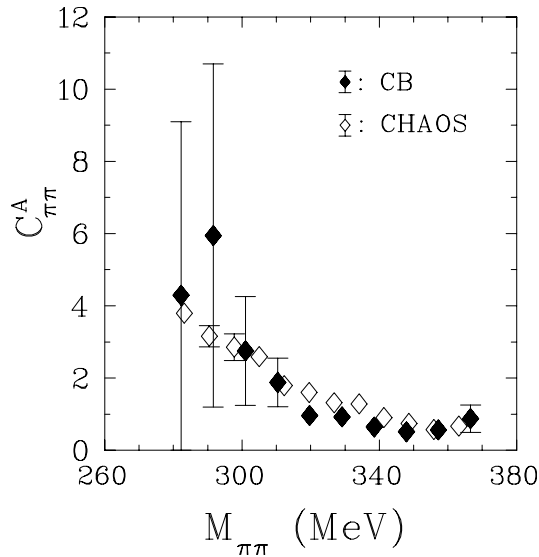}
 \vspace{0.5cm}
\caption{The composite ration as a function of the 
 $\pi$-$\pi$ invariant mass for the CB data (filled diamonds)
 and CHAOS data (open diamonds).
 The target is $^{12}$C in both cases. This figure is 
 taken from {\protect \cite{CAM01}} .}
    \label{fig11}
\end{figure}

For the experimental confirmation of 
 the threshold enhancement seen in \cite{CHAOS00},
 measurements of  2$\pi^0$ and 
$2\gamma$ final states with hadron/photon beams off
 the  heavy nuclear targets are necessary.
 Those channels are free from the $\rho$  meson background
  inherent in the $\pi^+\pi^-$ measurement.
 One of such  experiments measuring 
 2$\pi^0$ from the reaction
 $A(\pi^-, \pi^{0}\pi^{0})A'$ with 
 $p_{\pi^-}=408 $ MeV and 
 $A$= H, $^2$H, C, Al, Cu
 has been  done by the Crystal  Ball Collaboration
\cite{CB00}. They report
 some qualitative difference of the 
 $\pi$-$\pi$ invariant mass distribution from the 
 CHAOS data. For making a better comparison between the data
from different detectors with different acceptance,
the following ``combined  ratio'' is useful \cite{BON99},
\beq
C^{\bf A}_{\pi\pi}(M_{\pi\pi})=
\frac{\sigma^{\rm A}(M_{\pi\pi})/\sigma_{\rm T}^{\rm A }}
{\sigma^{\rm N}(M_{\pi\pi})/\sigma_{\rm T}^{\rm N }},
\eeq
where $\sigma_{\rm T}^{\rm A}\,(\sigma_{\rm T}^{\rm N})$
is the measured total cross section of the $(\pi, 2\pi)$ process
in nuclei (nucleon): This ratio represents the net effect of
nuclear matter on the interacting $(\pi\pi)_{I=J=0}$ system.
As shown in Fig.\ref{fig11}, it has recently been argued
 that both the CB data and the CHAOS data are consistent 
and gives an  enhancement
near the $2m_{\pi}$ threshold \cite{CAM01}, 
although the statistics in the CB data is relatively poor.
For more detailed discussions, see the contributions
to this Proceedings from CHAOS and CB groups.

We list some of the possible experiments 
 related to the in-medium $\sigma$  in the following.

\begin{itemize}

\item
Measuring of 2$\gamma$'s from the electro-magnetic decay of  
the $\sigma$ or $(\pi\pi)_{I=J=0}$ in nuclear matter,
  although the branching ratio is small,
 is an interesting alternatives to the 
 2$\pi$ measurement  because of the small final state
 interactions. In such an experiment,
 one needs to fight with large 
 background of photons mainly coming from $\pi^0$s.
 Nevertheless,  if the enhancement is prominent,
 there is a chance to find the signal.  

\item
 When $\sigma$ has a finite three momentum,
one can also detect dileptons  through the scalar-vector 
mixing in matter owing to the violation of 
the charge conjugation symmetry at finite density
\cite{KUNI95}; $\sigma \to \gamma^* \to e^+ e^-$.
 (See also, \cite{WOLF}.)  

\item
The inverse process $\gamma + A \rightarrow 2\pi + X$
can be also used to produce the $\sigma$
or $(\pi\pi)_{I=J=0}$ system by the electro-magnetic probes.
Such experiments with electro-magnetic probes
 have been planned and are being performed
in SPring-8\cite{SPRING8}.

\item
 We  remark that the nuclear reactions such as (d, $^3$He) and (d, t)
 are also useful to explore the
 spectral enhancement and/or the formation of the
 $\sigma$-meson nuclei \cite{HIR00} 
 as in the case of the deeply bound pionic nuclei
and the possible production of  $\eta$- or
$\omega$-mesic nuclei\cite{HHG99}.
The incident kinetic energy $E$ of the  deuteron in the laboratory
system is  estimated to be  $1.1 {\rm GeV} < E < 10$ GeV, 
 to cover the spectral function 
 in the range  $2m_{\pi} < \omega < 750$ MeV.

\end{itemize}

\subsection{Medium effect in other channels}

\begin{figure}[tbp]
   \centering
   \epsfxsize=7.0cm
   \epsfbox{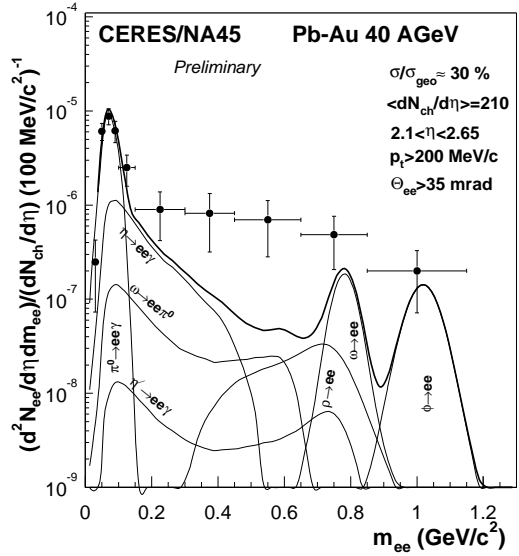}
 \vspace{0.5cm}
\caption{Normalized $e^+e^-$ invariant mass spectrum
 in the Pb-Au collision at 40 A GeV.
 Each contribution from the known hadronic sources are
 shown by thin solid lines, and their sum is shown
 by thick solid line. This figure is taken from the 
 second reference in  {\protect \cite{CERES01}}. }
    \label{fig12}
\end{figure}

\begin{figure}[tbp]
   \centering
   \epsfxsize=12.0cm
   \epsfbox{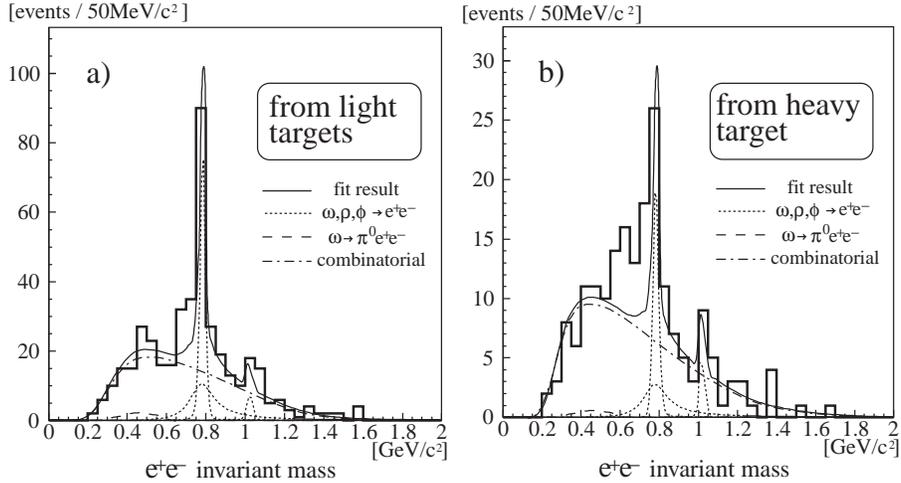}
  \vspace{0.5cm}
\caption{$e^+e^-$ invariant mass spectrum in the 
 $p$-$A$ reaction. (a) is for the C and polyethylene targets
 and (b) is for the Cu target. The solid lines in both
 figures are the result of the best-fit including
 known dilepton sources and the combinatorial background.
 Excess over the solid line is seen below the $\rho$-$\omega$
 peak in the case (b).}
    \label{fig13}
\end{figure}

The dilepton spectrum measured in the
heavy ion collisions such as Pb-Au collisions
 both at high energy (158 GeV/A) and at low energy (40 GeV/A)
 show a sizable enhancement of the $e^+e^-$ yield
 below the $\rho$-meson peak  \cite{CERES01}.
 This may or may not be 
 related to the partial chiral restoration in nuclear
medium originally proposed in \cite{P-BR-HL}
 (for the recent review, see \cite{RW01}).
 In Fig.\ref{fig12}, the recent CERES/NA45 data for 40 GeV/A collision are
shown. 
 The E325 experiment at KEK  \cite{OZAWA01}  measured
 $e^+e^-$ pairs from the $p$-A collision at 12 GeV.
 The similar enhancement over the known source and
 combinatorial background as CERES is seen in the mass range
 of about 200 MeV below
 the $\rho$-$\omega$ peak for $A=Cu$ (Fig.\ref{fig13}).

  The deeply bound pionic atom has proved 
to be a good probe of the properties of the hadronic interaction
 deep inside of heavy nuclei \cite{ITA98}.  It is  suggested that
the anomalous energy shift of the pionic atoms (pionic nuclei)
owing to the strong interaction could be attributed to the
 decrease of the effective pion decay constant 
$f^{\ast}_{\pi}(\rho)$ at finite density $\rho$ which 
may imply that the chiral symmetry is partially restored deep
inside the nuclei \cite{ITA98,WEISE00}.

\section{ Summary and concluding remarks}

This  report may be summarized as follows.

 The QCD phase diagram should be considered in 
  $m_i$-$T$-$n_B$ 
 space where $m_i$ ($i=u, d, s$) being the current 
 quark mass,  $T$ ($n_B$) is the temperature 
 (baryon density).
 the recent extensive studies based on  the 
 lattice QCD simulations and effective theories of QCD
 indeed  shows it quite probable that the chiral phase transition 
 takes place for $T_c \simeq 170 {\rm MeV}$ and $n_B \sim $
 several $\times$ $n_0$, although the precise order of the 
 transition in the real world is not clear yet.

The order parameter of the chiral symmetry breaking is
 $\langle \bar{q}q \rangle $. Furthermore,
   its amplitude and phase fluctuations
  are the ur$\sigma$ and the pion, respectively,
 which play a central
 role in hadronic world. They mix strongly through the
 process $\sigma \leftrightarrow 2 \pi$, which may account
 for the fact that  the physical $\sigma$ 
 is elusive in the real world.
 Irrespective of the introduction of the ur-$\sigma$, 
 it has become more and more  plausible in recent years that
 the strong $\pi$-$\pi$ attraction produces a light $\sigma$ 
 pole in the 2nd sheet of the $\pi$-$\pi$ scattering
 amplitude in the $I$=$J$=0 channel.
 Constraints from chiral symmetry, analyticity and crossing symmetry
 are key ingredients for the confirmation of the light $\sigma$-pole.
  Such a scalar-isoscalar fluctuation
 has also  phenomenological importance in modern hadron physics.

If  $T$ and/or $n_B$ of the system increases, 
 the intimate connection between the QCD vacuum structure and
 the fluctuation modes becomes  even more apparent.
 In fact,  the light $\sigma$ meson acts as a 
  soft mode and the complex $\sigma$-pole in the second sheet
 has characteristic shift toward the real axis in  association
 with the partial restoration of chiral symmetry
 $\langle \bar{q}q \rangle \rightarrow 0$.

 In cold nuclear matter,  a sizable decrease of the quark condensate,
 as large as 30 \% at normal nuclear matter density,
 may take place according to the
 low-density expansion of  $\langle \bar{q}q \rangle $.
 If this is the case, probing the scalar-isoscalar 
 fluctuation in cold nuclear matter using
 heavy nuclei is promising  for exploring
 the partial restoration of chiral symmetry.
 The softening of the complex $\sigma$-pole
  leads to a strong
 enhancement of the spectral function and the 
 $\pi$-$\pi$ cross section
 in the $I$=$J$=0 channel near the $2m_{\pi}$ threshold.
   This phenomena can be explicitly demonstrated e.g. in the
 linear and nonlinear realization of chiral symmetry
 provided that the
 possible reduction of the quark condensate or $f_{\pi}$
 is taken into account.

 The possible relevance of the above theoretical considerations
 and the $(\pi^{+}\pi^{-})_{I=J=0}$ data (the CHAOS collaboration) 
 and the $(\pi^{0}\pi^{0})$ data (the Crystal Ball collaboration)
 is  suggestive and  should be studied further.
 Independent tests using  the 
 electro-magnetic probes  as well as the mesic-nuclei
 will be also useful to unravel the realization of chiral symmetry
 in hadronic matter.

\vspace{.5cm}
This work is partially supported by Grants-in-Aid of the 
Japanese Ministry of Education Science and Culture (No. 12640263
 and 12640296).

\end{document}